\documentclass[aps,12pt,tightenlines]{revtex4}
\usepackage{graphicx,bm,epsf}

\begin{document}

\title{
    Possibility to study $\bm\eta$-mesic nuclei and photoproduction \\
    of slow $\bm\eta$-mesons at the GRAAL facility }

\author{
V.A. Baskov$^1$,
J.P. Bocquet$^2$,
V. Kouznetsov$^3$,
A. Lleres$^2$,
A.I. L'vov$^1$,\\
L.N. Pavlyuchenko$^1$,
V.V. Polyanski$^1$,
D. Rebrevend$^2$,
G.A. Sokol$^{1,}$\footnote{E-mail: gsokol@x4u.lebedev.ru}
}

%\address{
\affiliation{
$^1$P.N.~Lebedev Physical Institute, Leninsky prospect 53,
    119991 Moscow, Russia \\
$^2$Laboratoire de Physique Subatomique et de Cosmologie,\\
53, av.~des Martyrs, F-38026 Grenoble Cedex, France \\
$^3$Institute for Nuclear Research, 60-th October Anniversary prospect 7a,
   117312 Moscow, Russia
}

\begin{abstract}

A new experiment is proposed with the aim to study $\eta$-mesic nuclei
(bound systems of the $\eta$-meson and a nucleus) and
low-energy interactions of $\eta$ with nuclei.
Two decay modes of $\eta$ produced by a photon beam in the subprocess
$p (\gamma,p_1) \eta$ on a proton $p$ inside the nucleus $A$
will be observed, namely a collisional decay
$\eta N \to \pi N$ inside the nucleus and the radiative decay
$\eta \to \gamma \gamma$ outside. In addition, a collisional decay of
stopped $S_{11}(1535)$ resonance inside the nucleus,
$S_{11}(1535) N \to N N$, will be studied.
Triggers of the corresponding reactions are triple $(\pi^0 p p_1)$,
$(\gamma\gamma p_1)$, or $(p_1 p_2 p_3)$ coincidences
which include two particles from the $\eta$ or $S_{11}(1535)$ decays
and a nucleon $p_1$ flying forward in the case when
a slow $\eta$ is produced.

$\eta$-nuclei are expected to be observed as peaks in the energy
distribution of $\pi N$ or $NN$ pairs. This will bring an information
on binding energies and widths of $\eta$-nuclei, as well as an
information on $S_{11}(1535)$-nucleus interactions
at subthreshold energies of $\eta$.
Studies of slow-$\eta$ production through the radiative decay
mode will give an information on $\eta$-nucleus interaction
at energies above threshold of $\eta$ production.

The experiment can be performed using the tagged photon beam at ESRF
with the end-point energy $E_{\gamma\,\rm max} \approx 1000$ MeV
and the GRAAL detector which includes a high-resolution BGO calorimeter
and a large acceptance lead-scintillator time-of-flight wall.
Some results of simulation and estimates of yields are given.

\end{abstract}

\maketitle

\newpage
\tableofcontents

\newpage
\section{Introduction: $\bm{\eta N}$ interaction and $\bm\eta$-nuclei}
\subsection{Sources of information}

Interaction of $\eta$ mesons with other hadrons including nucleons and
nuclei is not yet well understood.  Being a member of the $SU(3)$ octet
of pseudoscalar mesons, $\eta$ has a spatial structure similar to that of
the pion.  However, $\eta$ has isospin 0 and
contains about 50\% of strange quarks ($s\bar s$).
It is much heavier than the pion (547 MeV versus 140 MeV).
Moreover, $\eta$ (and especially $\eta'$) is strongly mixed with gluons
which contribute through the $U_A(1)$ anomaly and make $\eta'$ heavier than
$\eta$.  For all these reasons, chiral symmetry which underlies
pion dynamics at low energies put far less constraints
onto the low-energy dynamics of $\eta$.

Nowadays, it is generally assumed that $\eta N$ interaction at low
energies is dominated by production and decay of the intermediate
$S_{11}(1535)$ resonance ($J^P = {\frac12}^-$)
which overlaps $\eta N$ threshold
($m_\eta + m_N = 1486$ MeV) within the total width of the
resonance, $\Gamma[S_{11}(1535)] \simeq 150$ MeV.
Most of experimental information on the $\eta N$ interaction comes from
studies of the reactions $\pi^- p \to \eta n$ and $\gamma N \to \eta N$
followed by their comparison with similar processes, in which $\eta$
is replaced by the pion (i.e.\ $\pi N \to \pi N$ and $\gamma N \to \pi N$).
Such a comparison is not straightforward and subject to model
uncertainties.

Assuming the $S_{11}(1535)$ dominance, the $s$-wave
Breit-Wigner cross sections of the above reactions are proportional
to partial widths of the $S_{11}(1535)$ resonance:
\begin{eqnarray}
     \sigma(\pi N\to\pi N) &\sim& \Gamma_\pi \Gamma_\pi,
\nonumber\\
     \sigma(\pi N\to\eta N) &\sim& \Gamma_\pi \Gamma_\eta,
\nonumber\\
     \sigma(\gamma N\to\pi N) &\sim& \Gamma_\gamma \Gamma_\pi,
\nonumber\\
     \sigma(\gamma N\to\eta N) &\sim& \Gamma_\gamma \Gamma_\eta.
\label{channels}
\end{eqnarray}
From the first two channels, the $\pi$- and $\eta$-widths and couplings
of $S_{11}(1535)$ can be found. Furthermore, the cross section of
$\eta N$ scattering can be derived too via
\begin{equation}
     \sigma(\eta N\to\eta N) \sim \Gamma_\eta \Gamma_\eta.
\label{eta-channel}
\end{equation}
Following this line in a modified form (in the framework of
the couple-channel formalism), Bhalerao and Liu \cite{bha85}
estimated the $\eta N$ scattering amplitude
and found, in particular, the $\eta N$ scattering length
\begin{equation}
        a_{\eta N} = (0.27 + i\, 0.22)~\rm fm.
\label{a-old}
\end{equation}
Using the third channel in (\ref{channels}), the radiative width of
$S_{11}(1535)$ can also be determined.
A result obtained by the VPI-GWU group \cite{arn96} and based
on their partial-wave analysis of pion photoproduction (SAID) reads
    $A_{1/2}^p \simeq  0.060~\rm GeV^{-1/2}$ (for the proton target)
and $A_{1/2}^n \simeq -0.020~\rm GeV^{-1/2}$ (for the neutron one)
in terms of the photocouplings $A_{1/2} \sim \sqrt{\Gamma_\gamma}$.
Alternatively, the same radiative widths and photocouplings
can be determined from the fourth channel in (\ref{channels}).
Using Mainz data on $\eta$ photoproduction off
protons and deuterons, the result was obtained \cite{kru95,roe96}
which was dramatically different from the previous findings:
    $A_{1/2}^p \simeq  0.110~\rm GeV^{-1/2}$
and $A_{1/2}^n \simeq -0.8 A_{1/2}^p$.
Thus, the radiative decay width of $S_{11}^+(1535)$, being determined
through $\pi$-photoproduction, is by the factor of $\sim 4$
less than the one determined through $\eta$-photoproduction.
This difference is even larger in the case of $S_{11}^0(1535)$.

Such a large discrepancy implies that the factorization
of the cross sections cannot be valid and therefore large nonresonance
backgrounds persist in some, if not all, considered channels.
In Ref.~\cite{chen02}, a dynamical model was developed, in which
nonresonance backgrounds were explicitly evaluated and
indeed found to be very essential and large.
It is often claimed that the radiative widths of $S_{11}(1535)$
must be determined through $\eta$ photoproduction, where the
background is allegedly small, and must not be determined through $\pi$ one,
where the background may be larger due to influence of a few
overlapping resonances.
In particular, in a recent update of the SAID analysis
of pion photoproduction \cite{arn02} the photocouplings of
the $S_{11}(1535)$ resonance are announced to be
''too uncertain to be quoted''.
Even so, our knowledge of $\eta N$ couplings and interactions
crucially depends on the factorization and extrapolation which leads us
from experimentally investigated channels (\ref{channels}) to the
unknown one, $\eta N$ scattering (\ref{eta-channel}).

In modern phenomenological analyses specific assumptions made
on resonances and backgrounds are different from those
assumed in \cite{bha85}.
In particular, more high-lying resonances are included,
what is effectively equivalent to considering the background
in the older approach energy dependent.
These changes, being physically marginal, have however a great impact
on the derived $\eta N$ scattering amplitude.
For example, the $\eta N$ scattering length
found in recent works through a coupled-channel analysis
of the first two reactions in (\ref{channels}) reads
\begin{eqnarray}
        a_{\eta N} &=& (0.75 \pm 0.04)
                 + i\, (0.27 \pm 0.03)~\rm fm \hspace{3em}\mbox{\cite{gre97}},
\nonumber \\
        a_{\eta N} &=& (0.717 \pm 0.030)
                 + i\, (0.263 \pm 0.025)~\rm fm \quad \mbox{\cite{bat97}}.
\label{a-new}
\end{eqnarray}
Note a dramatic difference in the real part of $a_{\eta N}$.

With more channels included and with the background structure
more complicated, the result for $a_{\eta N}$ may change again.
Currently, phenomenological analyses ignore channels
involving strange particles ($\pi N \to K \Lambda$, $\pi N \to K \Sigma$,
$\eta N \to K \Lambda$, and $\eta N \to K \Sigma$) ---
simply because of the lack of experimental data in the $\eta$ sector.
Meanwhile it was theoretically shown \cite{kai97} that an interplay
of strange channels with nonstrange ones can generate an energy-dependent
structure in $\pi N$ and $\eta N$ reactions which is very similar to that
usually associated with the $S_{11}(1535)$ resonance.
Thus, even the very dominance in the above reactions
of the fundamental three-quark state, the $S_{11}(1535)$ resonance,
over the dynamically generated coupled-channels peaks (hadron molecules)
becomes questionable.

In order to bring new experimental constraints, to check the
factorization, to test a truncated coupled-channel dynamics, and at the end
to arrive at a more reliable knowledge of properties of the $S_{11}(1535)$
resonance, direct data on $\eta N$ interaction are badly needed.
They can, in principle, be obtained through studies of $\eta$
in the final states of nuclear reactions.
This naturally suggests a need for 1) investigations of the
energy region near but above $\eta N$ threshold, where $\eta N$ scattering
affects cross sections through the final-state interaction,
and 2) investigations of the subthreshold energy region,
where formation of bound states of $\eta$ and a nucleus is possible.
Experimental studies of bound states of
various $\eta A$ systems would greatly contribute to learning
elementary $\eta N$ scattering and, more generally, to understanding
of meson-baryon interactions in the second nucleon-resonance region.

\begin{figure}[hbt]
%\centerline{\includegraphics[width=0.7\textwidth, bb=117 400 461 570]{p14p.ps}}
\centerline{\includegraphics[width=0.7\textwidth]{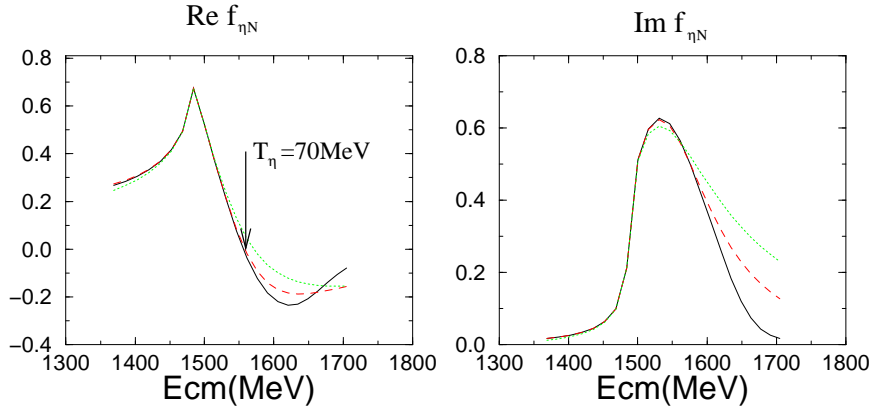}}
\caption{Energy dependence of the $\eta N$ scattering amplitude
    $f_{\eta N}$ (in units of fm) \protect\cite{gre97}.}
\label{f-etaN}
\end{figure}

\subsection{Eta-mesic nuclei}

Eta-mesic nuclei (${}_\eta A$) are a new kind of short-living nuclei
which consist of ordinary particles, i.e.\ nucleons, and a lighter one,
the $\eta$-meson, which is bound by nuclear forces with the nucleons.
The $\eta$-nuclei have many resemblances with better-studied $\Lambda$-
and $\Sigma$-hypernuclei which consist of nucleons and $\Lambda$ or
$\Sigma$, respectively.  A big difference is, however, in the life-time,
because $\eta$-nuclei decay via strong rather than weak interaction.
A typical width of bound-$\eta$ levels in nuclei is
of order $10{-}30$ MeV \cite{liu86,kul98,hai02}
or even $30{-}50$ MeV \cite{gar02,jid02}.
Annihilation of the bound $\eta$-meson through the reaction
$\eta N \to S_{11} \to \pi N$ in the nuclear matter dominates this width.
About $10{-}20$\% of the width is caused by the collisional
process $N S_{11}(1535) \to NN$ \cite{chi91}.
A tiny fraction of $\eta$-decays inside the nucleus
(of order $5 \cdot 10^{-5}$) happens due to the two-photon
decay $\eta \to \gamma \gamma$ too.
Due to virtual transitions between $\eta N$ states and the
$S_{11}(1535)$ resonance, part of time the $\eta$-nucleus is seen as a
complex of $S_{11}(1535)$ and $A-1$ nucleons.

The very existence of the bound $\eta$-A states \cite{pen85,liu86}
is possible because the positive real part of $a_{\eta N}$ results in
an average attraction between a slow $\eta$ and a nuclear matter.
This attraction is (roughly) characterized
by the first-order optical potential
\begin{equation}
     2m_\eta V_\eta(r) = - 4\pi \rho(r) a_{\eta N}
         \Big(1 + \frac{m_\eta}{m_N} \Big),
\end{equation}
where $\rho(r)$ is the nuclear density.
Depending on the magnitude of the scattering length $a_{\eta N}$,
such an attraction is sufficient for binding $\eta$ in all
nuclei of sufficient size --- namely those
with $A\ge 11$, if Eq.~(\ref{a-old}) is used,
or even those with $A\ge 4$, if newer values of $a_{\eta N}$,
Eq.~(\ref{a-new}), are valid. With slightly larger $a_{\eta N}$,
$\eta$-bound states are possible for $A=3$ and $A=2$ too \cite{rak96}.

When the kinetic energy of $\eta$ is not zero, the optical potential
becomes proportional to the $\eta N$ scattering amplitude $f_{\eta N}$.
The real part of $f_{\eta N}$, the threshold value of which is just
equal to the scattering length $a_{\eta N}$, is expected to remain
positive up to kinetic energies of $\eta$ below 70 MeV
(see Fig.~\ref{f-etaN}). This means that the effective $\eta A$ attraction
probably exists in a wide near-threshold energy region,
$\Delta E_{\eta} \approx 0{-}70$~MeV.

The attractive forces in the final state must lead to a
near-threshold enhancement in the total and differential cross section
of real-$\eta$ production by different beams. Such an enhancement was
indeed observed in several reactions including $p(d,{}^3\rm He)\eta$
\cite{ber88,wil97}, $d(d,{}^4\rm He)\eta$ \cite{may96} and
${}^2\rm H(\gamma,\eta),~{}^4\rm He(\gamma,\eta)$ \cite{hej02}, thus
supporting the existence of a rather strong ${\eta}A$ attraction even for the
lightest nuclei.  Nevertheless, all these experiments, which have deal
with $\eta$ in the final state, cannot directly prove that bound
$\eta{}A$ states do really exist.  A well-known counter-example is
provided by the $NN$ system in the $^1\rm S_0$ state, which has a
virtual rather than real level described by a {\em negative} rather
than positive scattering length.  Therefore, direct observations of
bound states through studies of the subthreshold energy region are
certainly needed in order to fully reconstruct the $\eta A$ optical
potential and the elementary $\eta N$ scattering amplitude. At energies
above the threshold, studies of real-$\eta$ production on nuclei with
different $A$ are needed in order to investigate the optical potential
at different nuclear densities.
Generally, experimental studies of various $\eta A$ systems
in the discrete and continuous spectrum would greatly contribute to
understanding of meson-baryon interactions
in the second nucleon-resonance region.

\subsection{Hunting $\eta$-mesic nuclei: negative results}

Two attempts to discover $\eta$-nuclei in the missing mass spectrum of
the reaction $\pi^+ A\to p X$ were performed at Brookhaven \cite{chr88}
and Los Alamos \cite{lie88} soon after the first theoretical
suggestions \cite{liu86}.
It was then expected that a capture of a produced slow $\eta$ to a
bound level(s) would result in a peak(s) of the width $\sim 10$ MeV
near the end of the kinetic energy spectrum of the knocked-out
forward-flying protons. However, narrow peaks were found neither in
\cite{chr88}, nor in \cite{lie88}.

An explanation was soon suggested in Ref.~\cite{chi91}, in which
the imaginary part of the optical $\eta A$ potential $V_\eta$
was argued to get a great increase owing to the
two-pion absorption of $\eta$ via the reaction
$\eta N \to S_{11}(1535) \to \pi\pi N$.
Accordingly, a great increase in the widths of bound-$\eta$ levels
was predicted too. The conclusion of Ref.~\cite{chi91} was that
the bound-$\eta$ levels in nuclei may be so wide that they strongly overlap
and hence a broad bump arises instead of a set of narrow lines
in the spectrum. This was exactly what the experiment \cite{lie88} showed,
in which a broad bump of the width $\ge 30$ MeV was observed approximately
at the place where narrow peaks were earlier predicted \cite{liu86}.

There is, however, a flaw in the above explanation, because
a large increase in Im\,$V_\eta$ is obtained in \cite{chi91} only with
the mass, the width and the couplings of $S_{11}(1535)$
which do not accurately reproduce available experimental data on the reaction
$\pi^-p\to\eta n$.  When these parameters are changed in order to
fit the data, the dramatic increase of the widths completely disappears
(see the second columns in the Table~1 of Ref.~\cite{chi91}).
In a more recent work \cite{kul98}, the absence of huge medium effects
on the imaginary part of the optical potential $V_\eta$ was re-established
and a physical picture with relatively narrow levels was confirmed.
Very recent calculations \cite{hai02,gar02,jid02} agree with these findings.

Still, there is another physical reason for unresolving narrow lines.
The point is that the $\eta$-mesic nucleus can be produced with excited
nucleon degrees of freedom. That is, the energy of the nuclear core
which binds $\eta$ is not fixed but rather smeared over the Fermi sea.
Such a Fermi smearing also contributes to broadening peaks in the
observed spectra and makes it more difficult to disentangle them
from a smooth background.

The conclusion is that the negative results \cite{chr88,lie88} of a
search for $\eta$-mesic nuclei in the previous missing-mass experiments
do not necessary mean that such physical objects either do not exist or
dissolve in the continuum. Rather the inclusive method of \cite{chr88,lie88}
was less reliable for searching for broad peaks and hence for measuring
energy levels of $\eta$-mesic nuclei than it was then assumed.

\subsection{Hunting $\eta$-mesic nuclei:
  positive results from photoproduction}

\begin{figure}[hbt]
\centerline{\includegraphics[width=0.8\textwidth]{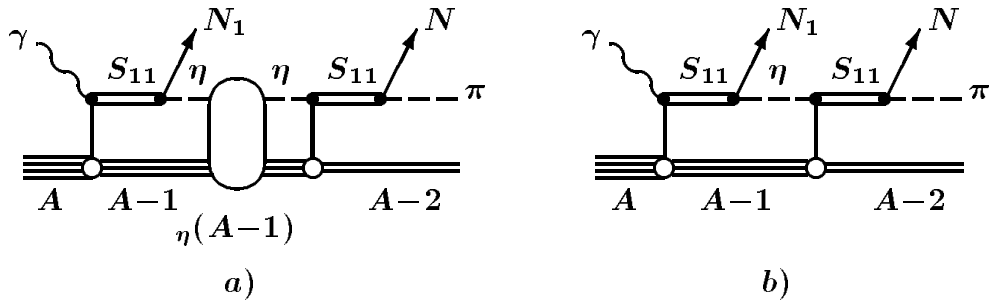}}
\caption{a) Mechanism of formation and decay of $\eta$-nuclei in
a photoreaction;
b) Production of $\pi N$ pairs through intermediate $\eta$
in the absence of $\eta A$ attraction.}
\label{mechanisms}
\end{figure}

Recently, the first positive signal for $\eta$-mesic nuclei was observed
in a photoreaction \cite{sok98}. In contrast to the inclusive
experiments \cite{chr88,lie88}, decay products of $\eta$-mesic nuclei
have been detected in that work.
The experiment was performed at the bremsstrahlung photon beam
of the 1~GeV electron synchrotron of Lebedev Institute.
The reaction studied was
\begin{equation}
 \gamma + {}^{12}{\rm C} \to p(n) + {}^{11}_{~\eta}{\rm B}~
    ({}^{11}_{~\eta}{\rm C}) \to p(n) + \pi^+ + n + X  = \pi^+ + n + X'.
\end{equation}
A slow $\eta$-meson produced in the subprocess
$\gamma N \to S_{11}(1535) \to \eta N$ is captured into a
quasi-bound state.  After some time-delay, it annihilates inside the nucleus
and gives a $\pi^+ n$ pair through the subprocess
$\eta p \to S_{11}^+(1535) \to \pi^+ n$ (see Fig.\ \ref{mechanisms}a).
A fast nucleon emitted forward at the first stage of the reaction
escapes the nucleus, whereas components of the $\pi^+ n$ pair fly
isotropically and they are detected by scintillator time-of-flight
spectrometers placed transversely to the photon beam in opposite directions.

\begin{figure}[hbt]
\epsfxsize=\textwidth
\epsfbox[68 568 558 728]{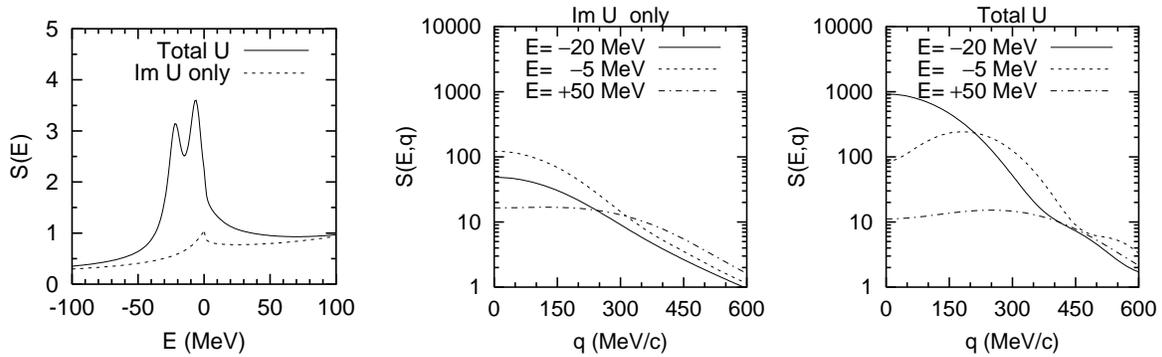}
\caption{Spectral functions $S(E)$ and $S(E,q)$ (in arbitrary units)
found with a rectangular-well optical potential
simulating the nucleus $^{12}$C.
For a comparison, shown also are results obtained with dropping out
the attractive (i.e.\ real) part of the $\eta A$ potential.}
\label{S(E)}
\end{figure}

Note that transversely flying $\pi^+ n$ pairs cannot be produced
via the one-step reaction $\gamma p\to\pi^+ n$ in the nucleus,
whereas they naturally appear through an intermediate slow-$\eta$ agent.
In principle, nothing prevents the chain
$\gamma N\to\eta N$, $\eta N\to\pi N$ to happen
in the absence of the $\eta$-nucleus attraction $V_\eta$ too
(see Fig.~\ref{mechanisms}b).
However, theoretical estimates given in \cite{lvo98,sok98} show that the
binding effects enhance the yield of the pairs with a low total 3-momentum
by the factor of $\sim 100$ and lead to a full dominance
of the reaction mechanism related with
a formation of the intermediate $\eta$-mesic nucleus
in the subthreshold region of the invariant mass
$\sqrt s_{\pi^+n} < m_\eta + m_N$. A clear peak in the invariant-mass
distribution is therefore predicted in the subthreshold region
(see Fig.\ \ref{S(E)} which show so-called spectral functions representing
probabilities to find an intermediate $\eta$-meson with the
(kinetic) energy $E$ and momentum $q$ \cite{lvo98,sok00}).

\begin{figure}[hbt] \unitlength=1mm
\begin{picture}(100,55)(0,0)
%\put(45,48){$E_{\gamma\,\rm max}=850$ MeV}
%\put(100,48){$E_{\gamma\,\rm max}=650$ MeV}
\put(10,48){$E_{\gamma\,\rm max}=850$ MeV}
\put(65,48){$E_{\gamma\,\rm max}=650$ MeV}
%\centerline{
\includegraphics[width=4.5cm, bb=157 322 443 613]{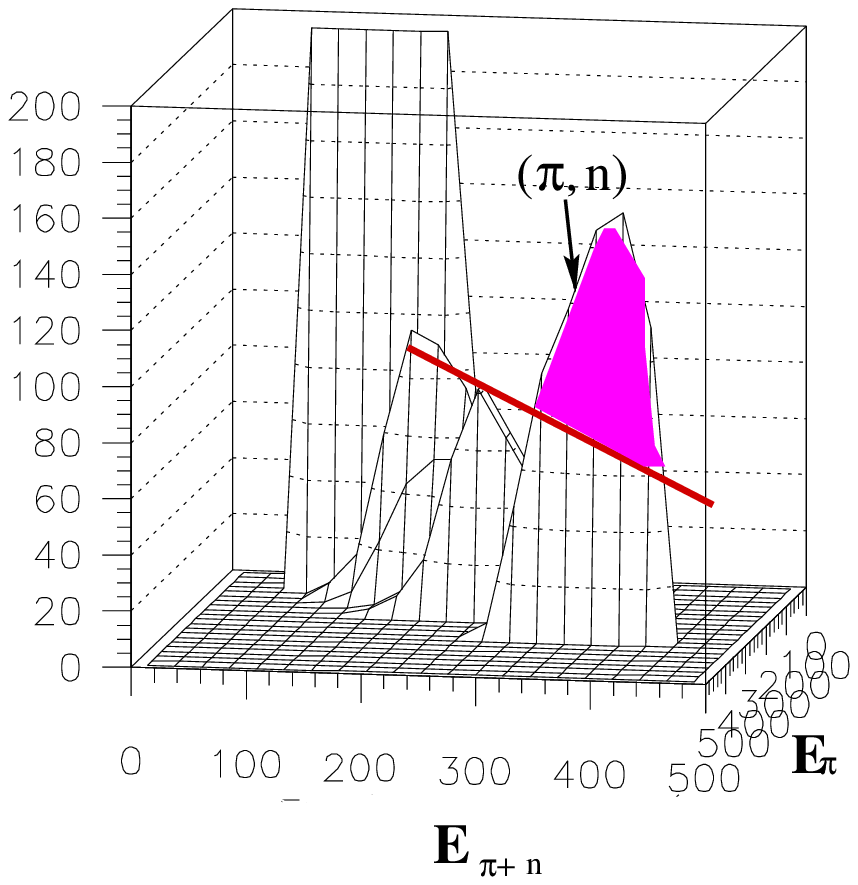}
\includegraphics[width=4.0cm, bb= 16 325 304 670]{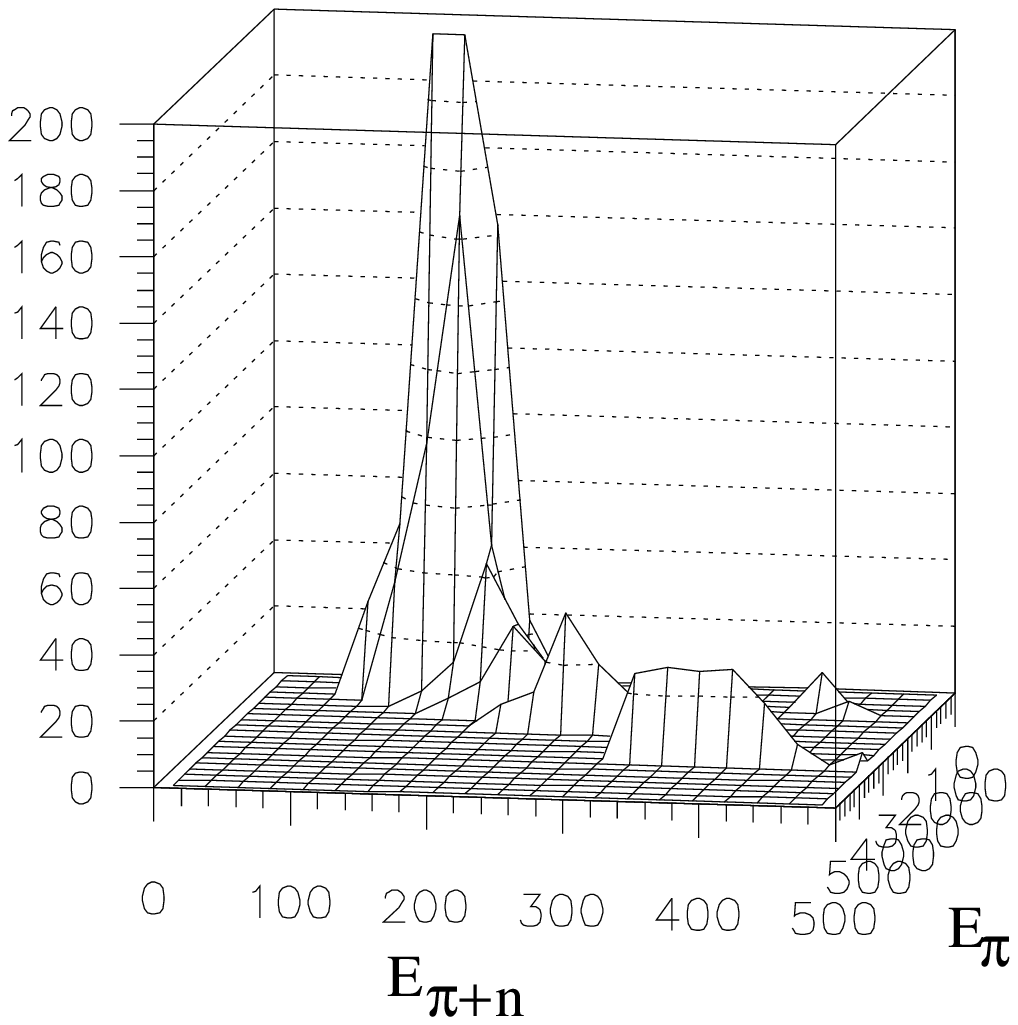}
%}
\end{picture}
\caption{Two-dimensional distributions over the total kinetic energy
of the $\pi^+n$ pair, $E_{\pi^+n}$, and
the kinetic energy of the $\pi^+$-meson, $E_{\pi}$,
for two energies $E_{\gamma\,\rm max}$ of the bremsstrahlung photon beam
\protect\cite{sok00}.}
\label{2dim}
\end{figure}

Such a peak was indeed observed in the experiment, provided
the photon beam energy $E_{\gamma\,\rm max}$
exceeded the $\eta$-production threshold off the free nucleon (707 MeV).
The peak was absent when $E_{\gamma\,\rm max}$ was below threshold.
See Fig.~\ref{2dim}, in which data shown are obtained by applying
an unfolding procedure to raw experimental velocity-distributions
found by the time-of-flight technique.
The unfolding procedure used is based on the statistical method of solving
the inverse problem \cite{pav82}.

\begin{figure}[hbt]
%\vspace{4mm}
%\centerline{\includegraphics[width=0.4\textwidth, bb=10 390 300 680]{1535n.ps}}
\centerline{\includegraphics[width=0.35\textwidth]{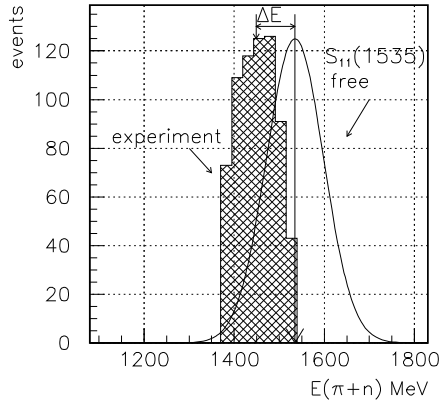}}
\caption{Distribution over the total energy of the $\pi^+n$ pair
after a subtraction of the background.}
\label{1dim}
\end{figure}

\begin{figure}[hbt]
\centerline{\includegraphics*[width=0.45\textwidth]{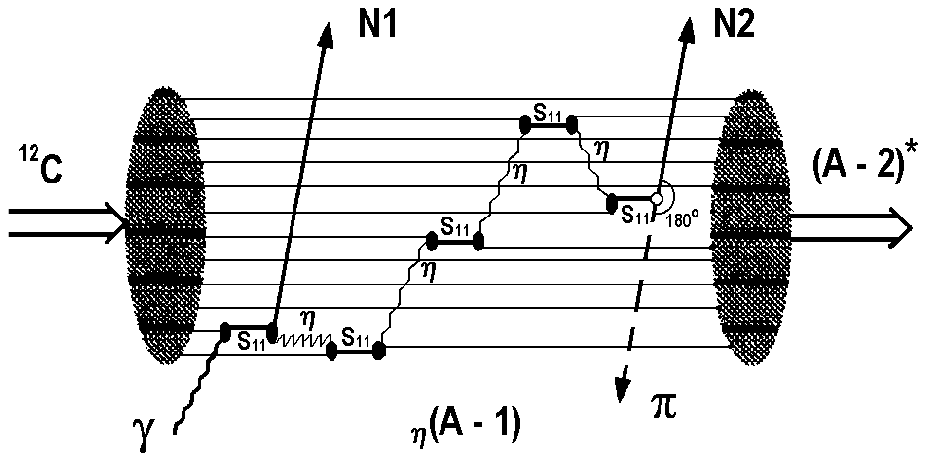}}
\caption{Photoproduction of a $\eta$-nucleus
and its decay through the $\pi N$ mode.}
\label{reac1}
\centerline{\includegraphics*[width=0.45\textwidth]{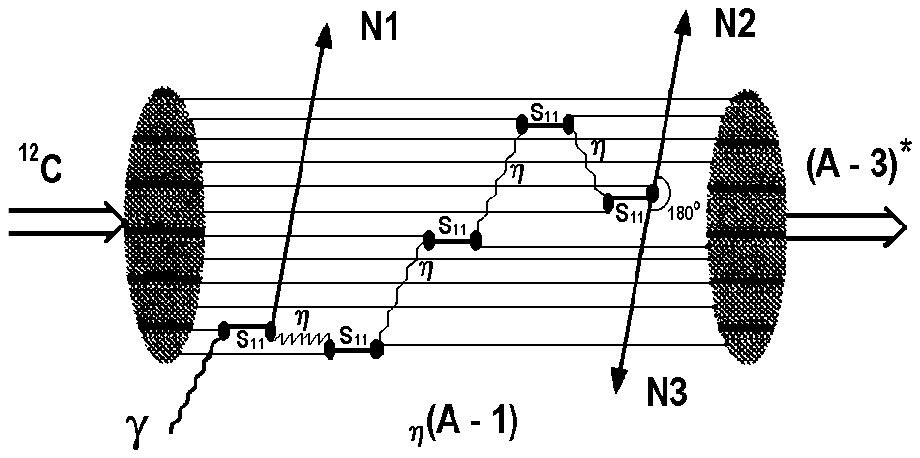}}
\caption{Photoproduction of a $\eta$-nucleus
and its decay through the $NN$ mode.}
\label{reac2}
%\centerline{\includegraphics*{fig8-3.eps}}
\centerline{\includegraphics*[width=0.4\textwidth]{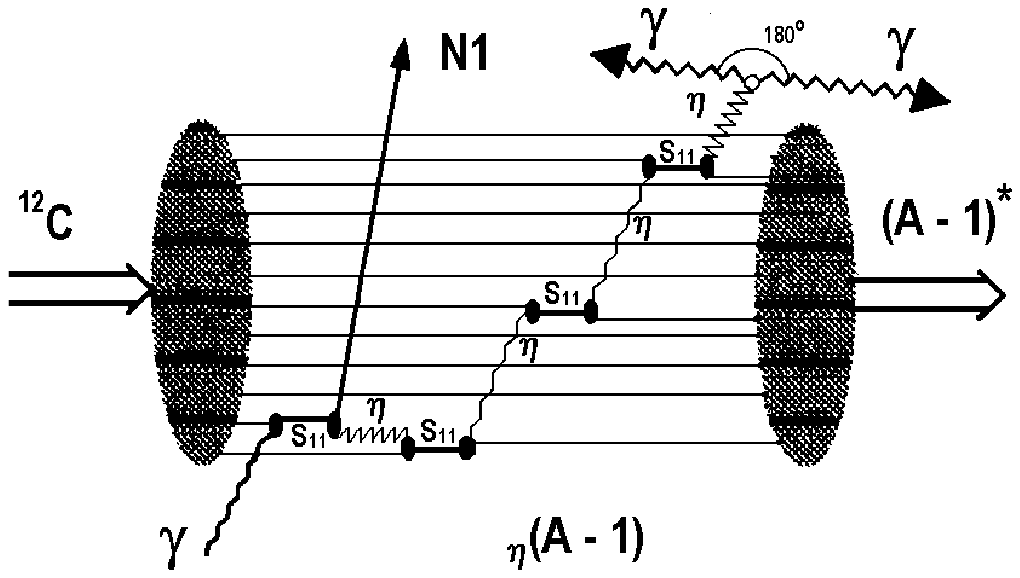}}
\caption{Photoproduction of a slow $\eta$-meson
and its decay through the $\gamma\gamma$ mode. This decay may happen
both inside and (mostly) outside the nucleus.}
\label{reac3}
\end{figure}

Subtracting a smooth background, a one-dimensional distribution of the
$\pi^+n$ pairs over their total energy has also been found (Fig.~\ref{1dim}).
Its width is about 150 MeV including the experimental resolution
of the setup.  Its center lies by $\Delta E=40 \pm 15$ MeV lower than
$\eta N$ threshold, $m_\eta+m_N=1486$ MeV, and by 90 MeV lower than the
$S_{11}(1535)$ resonance mass. The shift in the peak position
with respect to the threshold provides an evidence
for seeing binding effects for $\eta$ in the nucleus.

\section{Layout of a new experiment}

\subsection{The experimental setup and scenarios of measurements}

\begin{figure}[hbt]
\centerline{\includegraphics{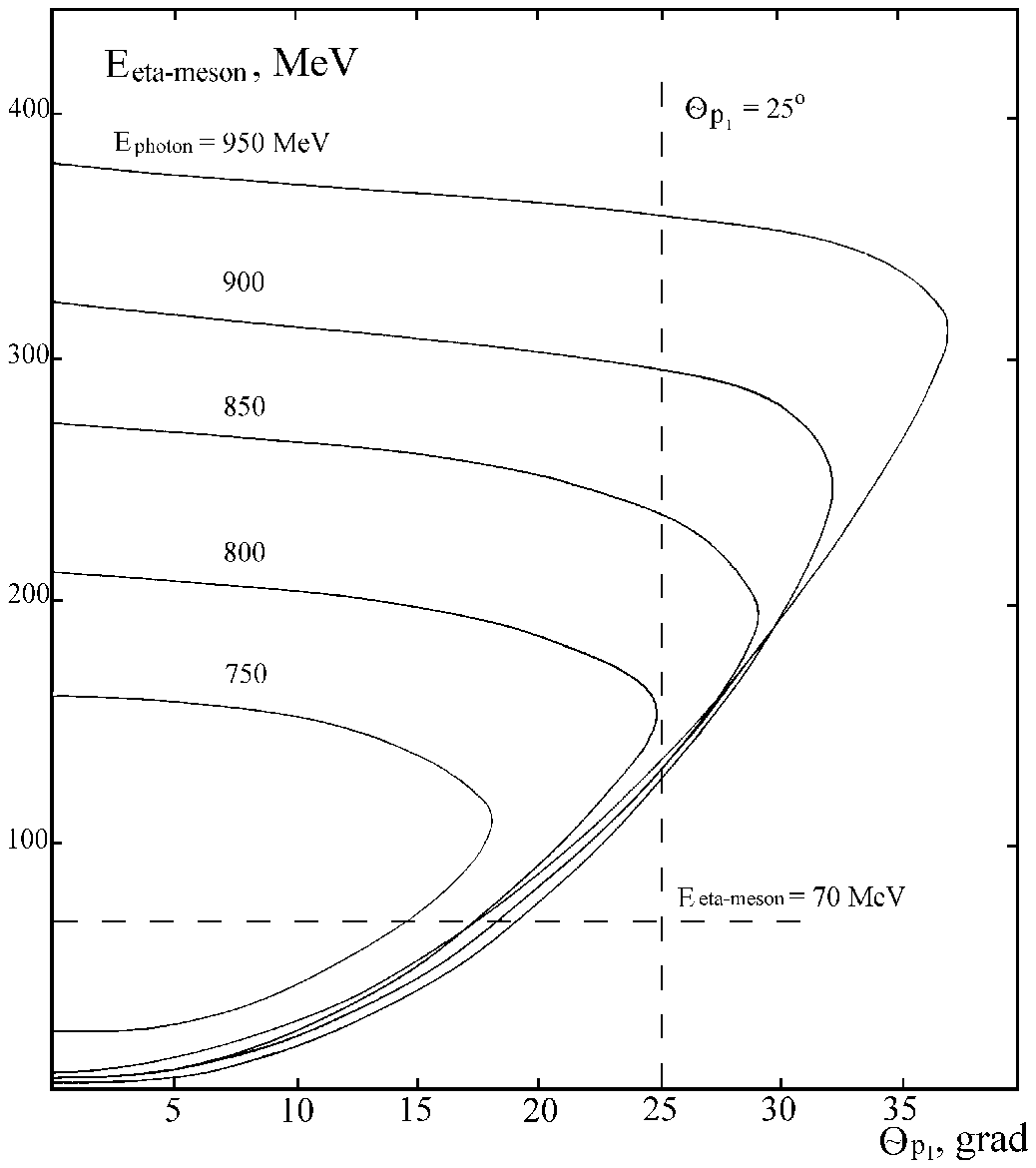}}
\vspace*{2cm}
\caption{Kinematics of $\eta$-meson photoproduction off the free proton,
$\gamma p\to\eta p_1$.}
\label{kin-p1}
\end{figure}

The main aim of the proposed experiment is measuring the energy-
and $A$-dependence of the cross section of $\eta$-mesic-nucleus
photoproduction, as well as measuring the ratio of $\pi N$ and $NN$ yields
which are connected with probabilities of the processes $S_{11}(1535)\to\pi+N$
and $S_{11}(1535)+N\to N+N$ inside different nuclei.
Also, data on production of slow $\eta$ off different nuclei
will be collected with the end aim to learn the optical potential $V_\eta$
in the near-threshold region. The reactions to be studied are
\begin{equation}
   \gamma + A \to p_1 + {}_\eta(A-1) \to p_1 + \pi^0 + p + X
   \to p_1 + \gamma_1 + \gamma_2 + p + X
\label{eq:reac1}
\end{equation}
(see Fig.~\ref{reac1}),
\begin{equation}
   \gamma + A \to p_1 + {}_\eta(A-1) \to p_1 + p_2 + p_3 + X
\label{eq:reac2}
\end{equation}
(see Fig.~\ref{reac2}), and
\begin{equation}
   \gamma + A \to p_1 + {}_\eta(A - 1) \to p_1 + \gamma_1 + \gamma_2 + X
\label{eq:reac3}
\end{equation}
(see Fig.~\ref{reac3}),
where the proton $p_1$ flying forward is produced at the stage of
creating a slow $\eta$-meson inside the nucleus via the subprocess
$\gamma p\to \eta p_1$ (the kinematics of this subprocess in the case of
$p$ at rest is shown in Fig.~\ref{kin-p1}). Depending on whether the
produced $\eta$-meson annihilates electromagnetically or
through the strong reaction $\eta p\to S_{11}(1535)\to\pi^0 N$,
either two photons with the invariant mass of the $\eta$-meson or two photons
from $\pi^0$ plus a proton flying in the opposite direction are to be
detected. In the case when the $S_{11}(1535)$ resonance annihilates
collisionally through the two-nucleon mode, two protons emerge which fly
in opposite directions. A rough estimate for relative rates of these three
outcomes (\ref{reac1}), (\ref{reac2}), (\ref{reac3})
is $1:10^{-1}:10^{-5}$.

In contrast to the first two cases (\ref{reac1}) and (\ref{reac2}),
the reaction (\ref{reac3}) directly provides, through the full energy
$E(\gamma_1+\gamma_2)$ of gammas, the energy of the $\eta$ produced.
Unfortunately, it is very difficult to use this energy information for
determination of the binding energy of $\eta$ in the nucleus because
of the very low probability $(\sim 10^{-5})$ of bound $\eta$ to decay
through the electromagnetic mode. Moreover, owing to a limited energy
resolution of photon detectors, the $\gamma\gamma$ events
from the bound-$\eta$ decays can be lost on the background of
$\gamma\gamma$ pairs arising after photoproduction of slow $\eta$-mesons
without a formation of the $\eta$-nucleus, what exactly happens
at energies slightly above threshold. Nevertheless, a measurement
of $E(\gamma_1+\gamma_2)$ provides a good way for studying energy dependence
of near-threshold $\eta A$ interaction.

%It is possible to select these cases for detection $\gamma \gamma$ events for
%decay $\eta$-meson inside and out nuclei of use diference in $\Sigma(\gamma_1 +
%\gamma_2)$ and $\Theta(\gamma_1 \gamma_2)$ for these reactions. For decay
%$\eta$-meson inside $\eta$-nuclei we have $\Theta_{\gamma_1 \gamma_2} = 180^0$
%and $E(\gamma_1 + \gamma_2) = m^* (\eta)$ but decay $\eta$-meson in free space
%we have $\Theta_{\gamma_1 \gamma_2} < 180^0$ for mainly cascs decay and
%$E_{\gamma_1 \gamma_2}$ in crise with energy $E\gamma()(E(\eta))$.
%It is need simulation of this two processes. We hope that relation of this two
%yeilds will be better than $1:10^{-5}$, we wait that this relation my be
%$1:10^3$.
%Using the tagged photons and detecting and measuring
%the energy of the proton knocked out in the process of quasi-free
%$\eta$-production in the nucleus, one can tag the energy of the $\eta$
%staying in the nucleus, $E_{\eta} \simeq E_\gamma - E_p^{\rm kin}$.
%Thus, studies with tagged $\eta$-mesons become possible what opens
%additional possibilities for learning the energy dependence of
%interactions between $\eta$ and nuclear constituents.

The experiment can be performed using the tagged photon beam at ESRF
with the maximum energy $E_{\gamma\,\rm max} \simeq 1000$ MeV and the GRAAL
apparatus. The trigger consists of a simultaneous detection of the
forward-flying proton $p_1$ plus two photons from $\pi^0$ or $\eta$.
In addition, the presence of a fast proton $p$ from a $\pi^0p$ pair
(decay products of $S_{11}(1535)\to\pi N$) is analyzed, as well as
the presence of two protons from the reaction $S_{11}(1535)+N\to N+N$.

\begin{figure}[hbt]
\centerline{\includegraphics*{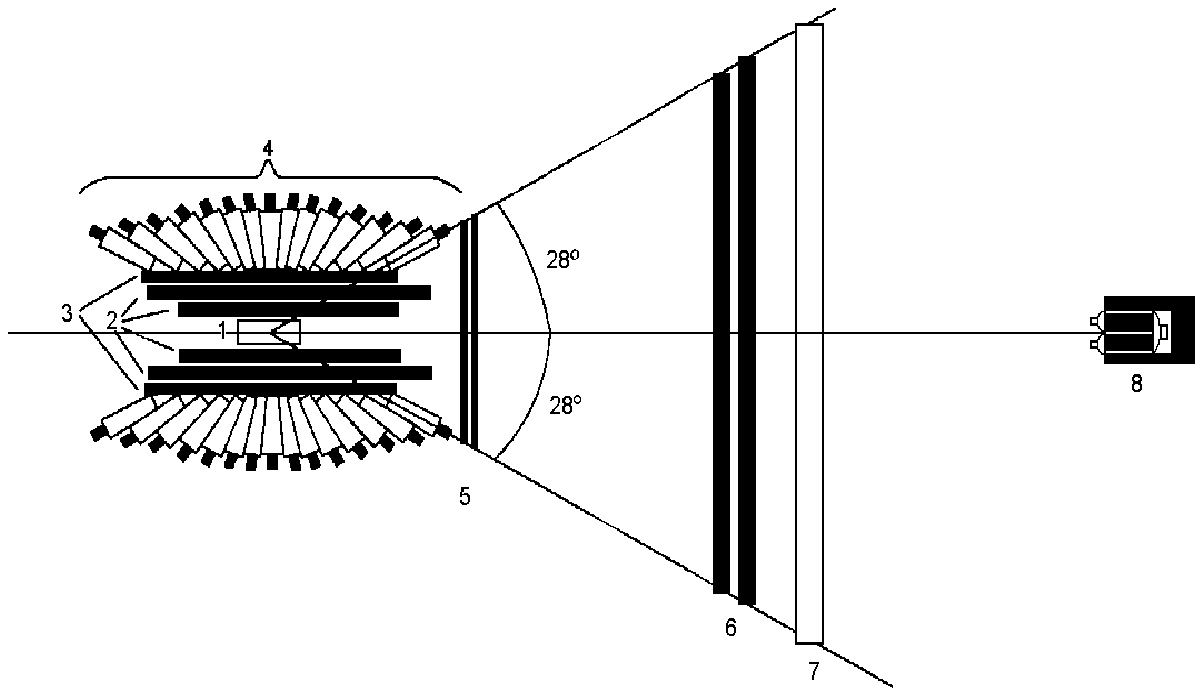}}
\vspace*{1cm}
\caption{Layout of the GRAAL apparatus.
 1: Liquid-hydrogen target;
 2: Cylindric wire chambers;
 3: Plastic scintillator barrel;
 4: BGO crystals;
 5: Plane wire chambers;
 6: Double scintillator wall;
 7: Sandwich calorimeter;
 8: Beam monitor.}
\label{graal}
\end{figure}

The GRAAL apparatus (Fig.~\ref{graal}) consists of a high-resolution
large solid-angle electromagnetic BGO-calorimeter combined with
multiwire proportional chambers (MWPC) which cover a solid-angle range
of near $4\pi$.
Particles emitted at small angles are also detected by
a scintillator wall which is installed three meter apart from the target;
it enables particle's identification by means of their time-of-flight and
their energy losses $\Delta E$ in scintillators.

The particle identification in the central region is accomplished with
a plastic scintillator barrel through the measurement of $dE/dx$.  The
BGO crystals are of 8 different dimensions. They are shaped like
pyramidal sectors with a trapezoidal basis. They define 15 angular
regions in the polar angle $\theta$ (the polar axis being the symmetry
axis of the calorimeter coincident with the photon-beam axis) and 32
angular regions in the azimuthal angle $\phi$.

All the 480 crystals have the same length of 24 cm
(i.e.\ $>21$ radiation lengths) ensuring a good absorption of photon
showers in the 1 GeV energy region. The crystals are arranged in such a way
that reaction products emitted in any directions around the target center
encounter a constant thickness of the BGO. More detail on the performance of
the BGO calorimeter and the time-of-flight wall can be found in
Refs.~\cite{ghi97,kou01}

In the experiment, the forward-flying proton $p_1$ will be detected by
the time-of-flight wall, whereas two photons from $\pi^0$ or $\eta$
decays plus protons from $\pi^0 p$ (with $E_p\sim 100$ MeV)
or $pp$ pairs (with $E_p\sim 300$ MeV) which are products of
$S_{11}(1535)\to\pi N$ and $S_{11}(1535)+N\to N+N$, respectively,
will be detected by the BGO calorimeter and multiwire chambers.

%In principle, others modes of the $S_{11}^+(1535)$
%and $S_{11}^0(1535)$ decays, $\pi^+ n$, $\pi^- p$ and $\pi^0 n$, are
%also observable. Moreover, in $10{-}20\%$ cases,
%the $S_{11}(1535)$ decays through collisions with nucleon,
%$S_{11}(1535) N \to NN$, where the final particles, i.e.\ fast
%($\sim 300$ MeV) $pp$, $pn$ or $nn$ pairs, can also be detected.

\subsection{Monte-Carlo simulation of the detection
 of $\pi N$ and $NN$ events}

\begin{figure}[hbt]
\centerline{\includegraphics*[height=8cm]{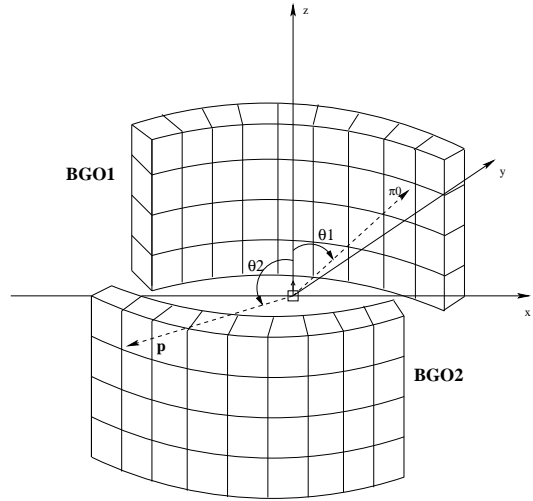}}
\caption{A simplified model of the BGO calorimeter used for a test
simulation of the detection of $\pi N$ and $NN$ pairs.
The photon beam is along the axis $z$.}
\label{BGOmodel}
\end{figure}

Since the BGO calorimeter does not ensure a sufficient
acccuracy in measuring energies of neutrons and charged pions, most
of $\pi N$ and $NN$ channels are not suitable for the planned measurements.
The exceptions are $\pi^0 p$ and $pp$ channels.
This statement is illustrated by a simple Monte-Carlo estimate using
the GEANT code. $10^3$ events were generated assuming a sample kinematics
in which the nucleon and the $\pi$-meson fly from the center of a small
$\rm 1~cm \times 1~cm \times 1~cm$ $^{12}$C target with the kinetic
energy of 100 MeV and 300 MeV, respectively, in opposite directions and
exactly transversely to the photon beam ($\theta_N=\theta_\pi=90^\circ$).
The BGO blocks were taken to have a semi-cylinder form each with the size
$\rm 50~cm \times 50~cm \times 25~cm$, so that the pion and the nucleon
strike the centers of the BGO1 and BGO2 blocks
shown in Fig.\ \ref{BGOmodel}).
Energy losses in the target and in the air between the target
and the calorimeter have been taken into account.

\begin{figure}[hbt]
\centerline{\includegraphics[height=7cm]{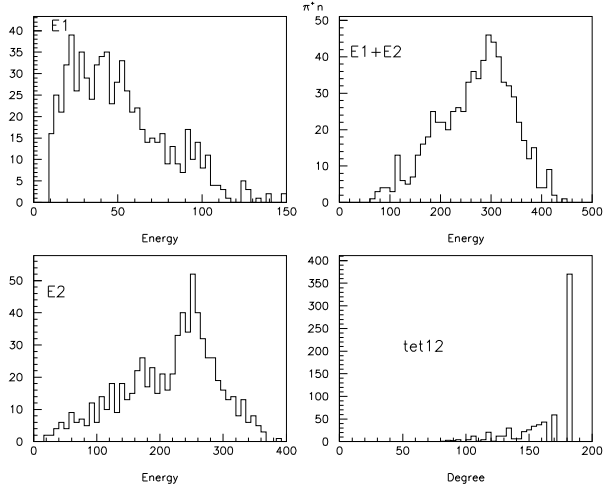}}
\caption{Energy depositions and angular distribution of $n$ and $\pi^+$
detected in the BGO calorimeter. Notation: $E1 = E_n$, $E2 = E_{\pi^+}$,
${\rm tet12} = \theta_{n \pi^+}$.}
\label{MC-n-pi+}
\end{figure}

\begin{figure}[hbt]
\centerline{\includegraphics[height=7cm]{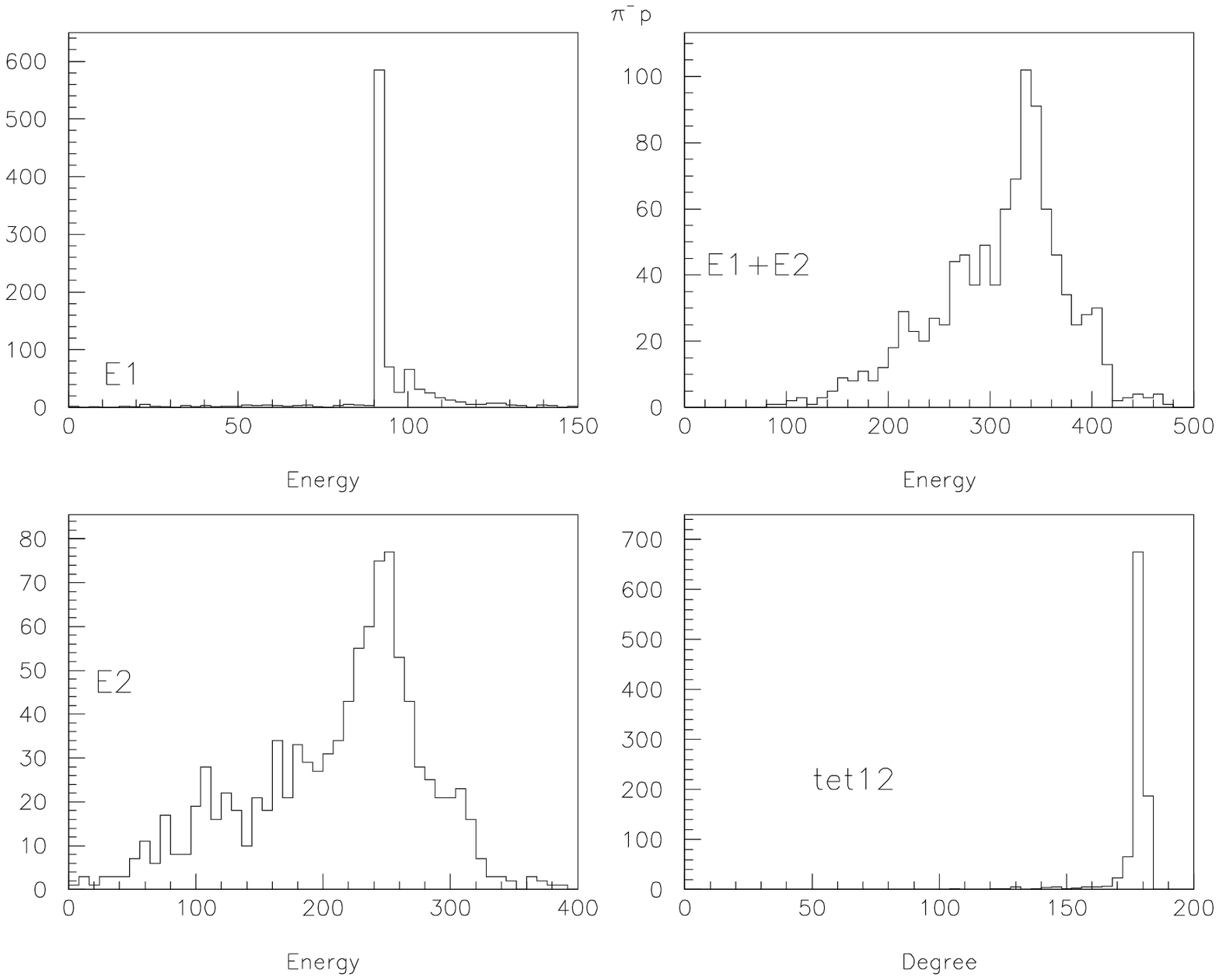}}
\caption{Energy depositions and angular distribution of $p$ and $\pi^-$
detected in the BGO calorimeter. Notation: $E1 = E_p$, $E2 = E_{\pi^-}$,
${\rm tet12} = \theta_{p \pi^-}$.}
\label{MC-p-pi-}
\end{figure}

\begin{figure}[hbt]
\centerline{\includegraphics[width=0.6\textwidth]{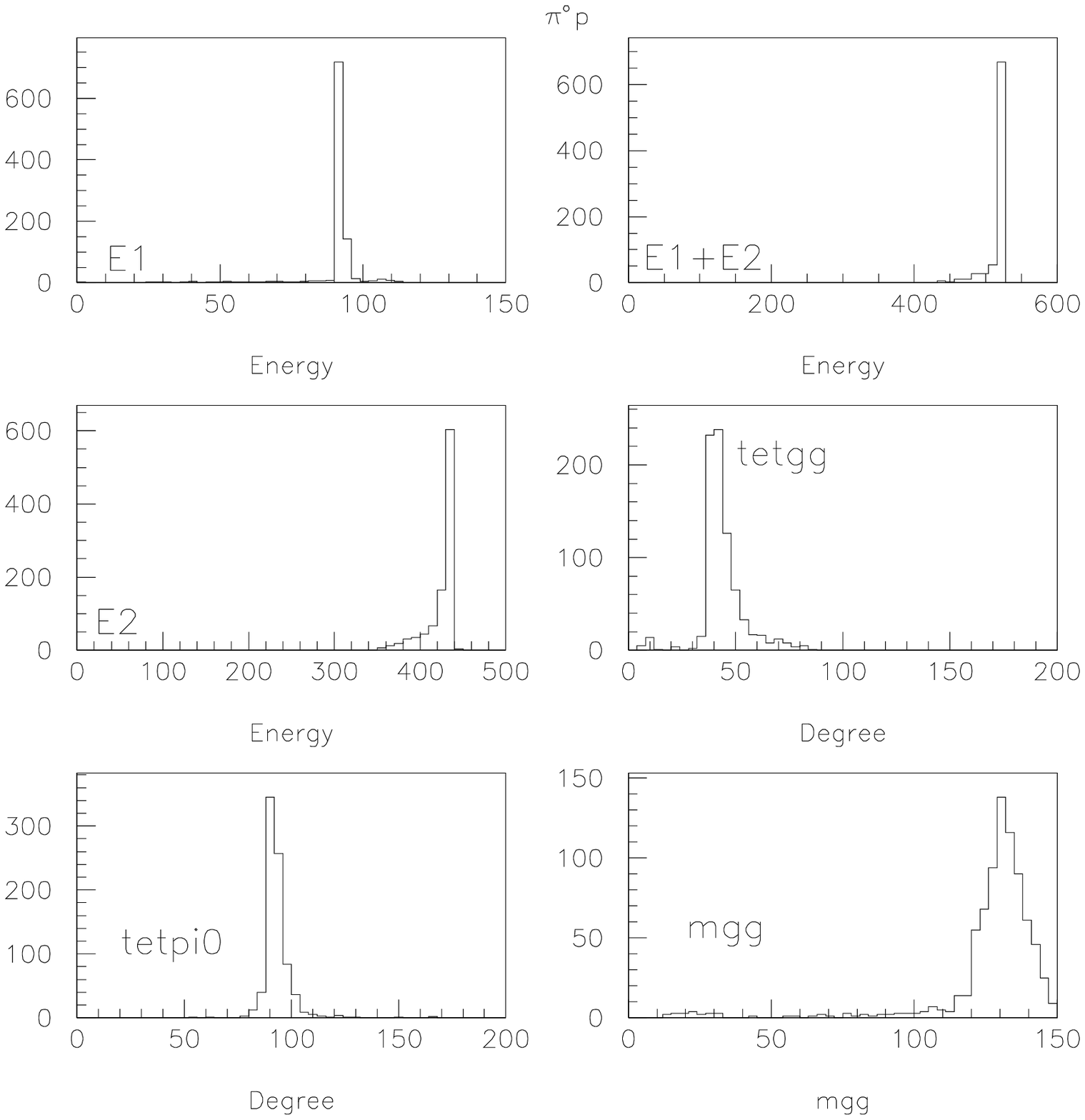}}
\caption{Energy depositions and angular distribution of $p$ and $\pi^0$
detected in the BGO calorimeter. Notation: $E1 = E_p$, $E2 = E_{\pi^0}$,
${\rm tetgg} = \theta_{\gamma_1\gamma_2}$,
${\rm tetpi0}$ is the reconstructed angle $\theta_{\pi^0}$, and
${\rm mgg}$ is the reconstructed invariant mass $m_{\gamma_1\gamma_2}$.}
\label{MC-p-pi0}
\end{figure}

\begin{figure}[hbt]
\centerline{\includegraphics[width=0.6\textwidth]{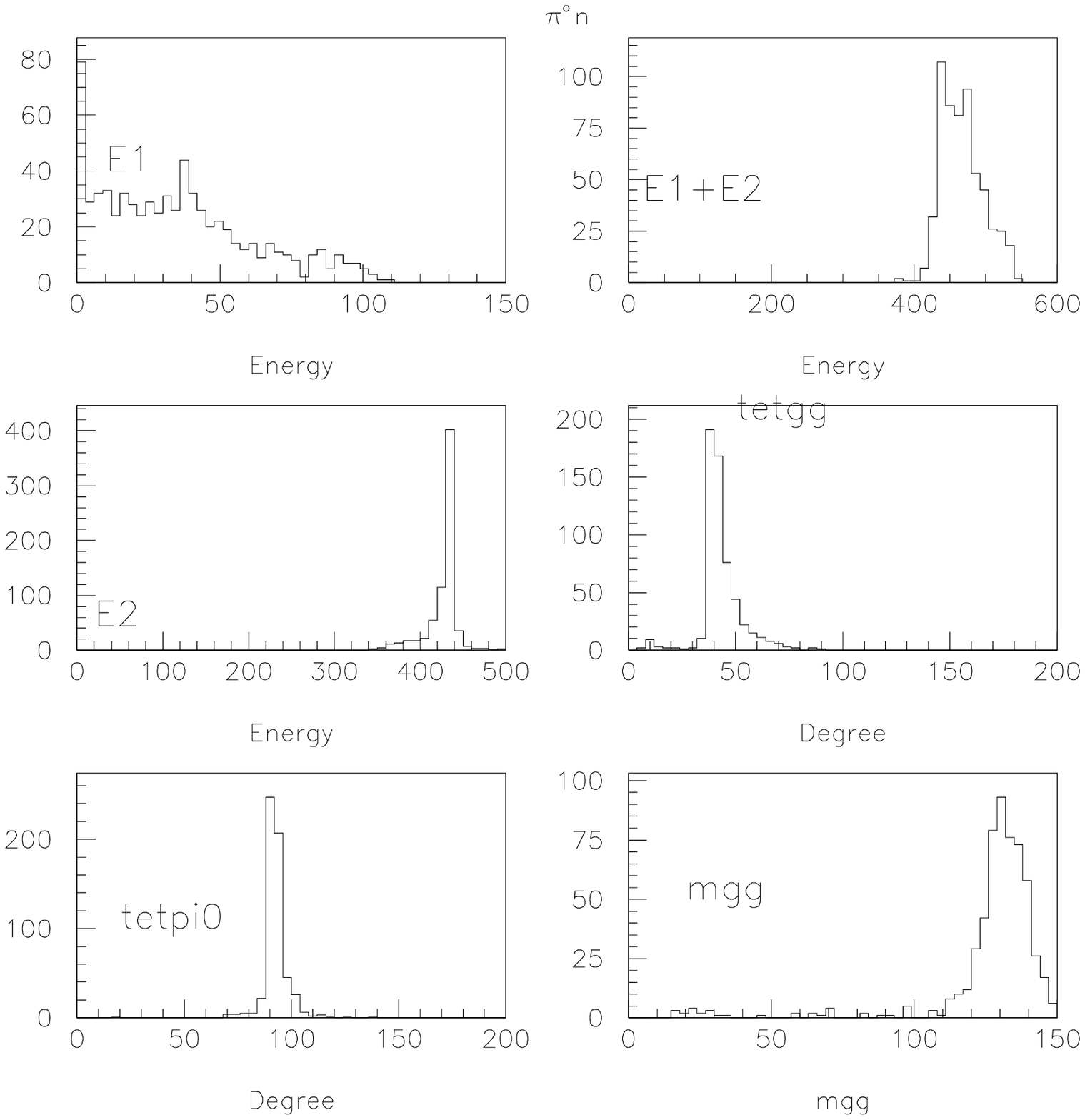}}
\caption{Energy depositions and angular distribution of $n$ and $\pi^0$
detected in the BGO calorimeter. Notation: $E1 = E_n$, $E2 = E_{\pi^0}$,
${\rm tetgg} = \theta_{\gamma_1\gamma_2}$,
${\rm tetpi0}$ is the reconstructed angle $\theta_{\pi^0}$, and
${\rm mgg}$ is the reconstructed invariant mass $m_{\gamma_1\gamma_2}$.}
\label{MC-n-pi0}
\end{figure}

Energy depositions in and angular distributions reconstructed through
such a model BGO calorimeter are shown in Figs.\ \ref{MC-n-pi+},
\ref{MC-p-pi-}, \ref{MC-p-pi0} and \ref{MC-n-pi0}
for the channels $\pi^+n$, $\pi^-p$, $\pi^0p$, and $\pi^0n$,
respectively. In the last two cases we also show a reconstructed direction
$\theta_{\pi^0}$ of the $\pi^0$ momentum and the reconstructed mass of
the pion,
$m_{\pi^0}=\sqrt{2 E_{\gamma_1} E_{\gamma_2}
    (1-\cos\theta_{\gamma_1\gamma_2})}$;
the deviation of these quantities from $90^\circ$ and 135 MeV, respectively,
gives a rough idea on the accuracy of using the BGO calorimeter for
reconstructing the events.

\begin{figure}[hbt]
\centerline{\includegraphics[height=8cm]{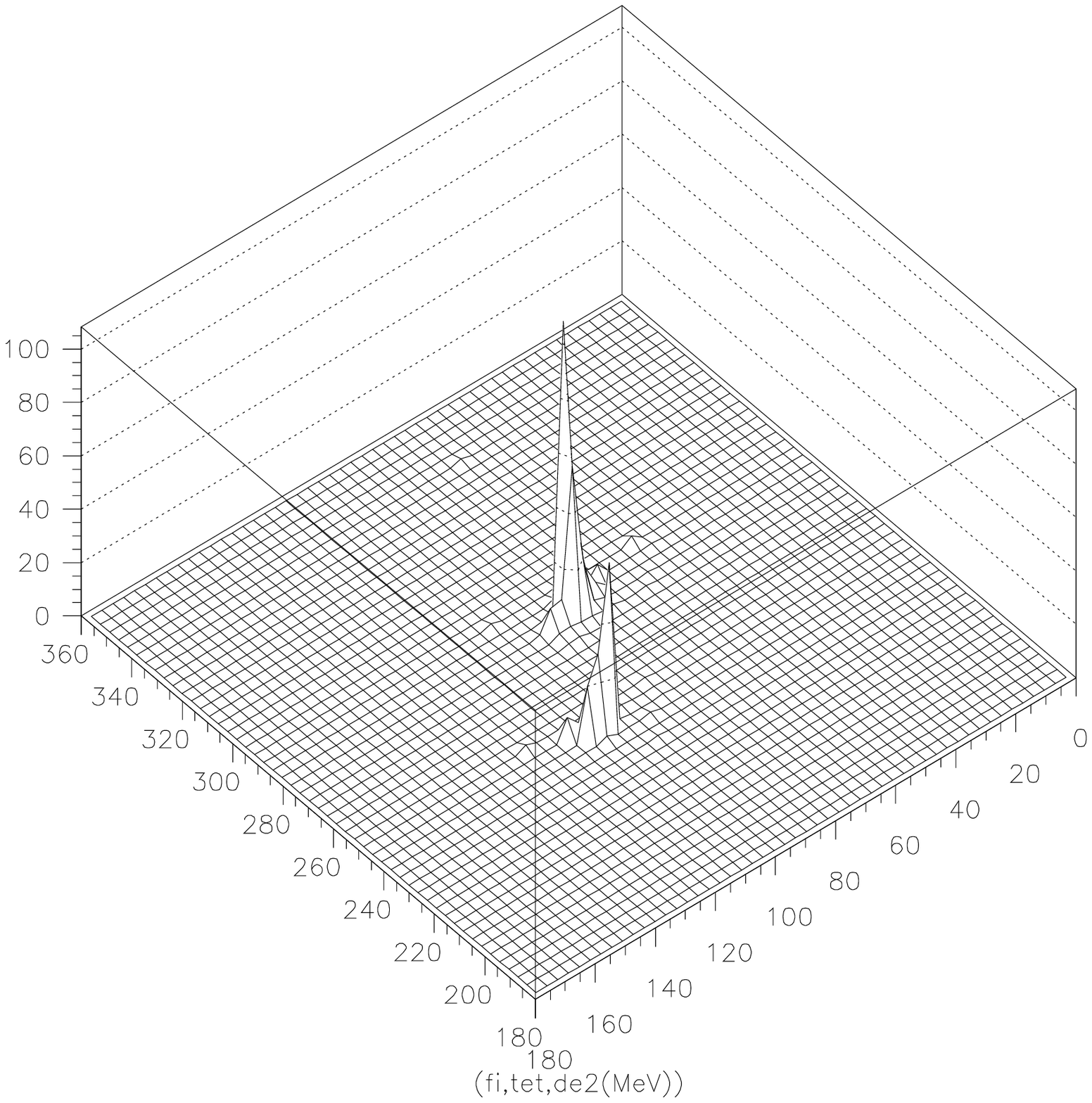}}
\vspace{4em}
\caption{The two-dimensional distribution of energy depositions
over the polar and azimuthal angles, $\theta$ and $\phi$,
for two photons, decay products of $\pi^0$ flying from the taget
at $(90^\circ,270^\circ)$ with the energy  $E_{\pi^0} = 300$ MeV.}
\label{MC2-p-pi0}
\end{figure}

Note, in particular, that the measured total energy $E_{\pi^0} + E_p$
gives us an estimate of the binding effect for $\eta$ in the nucleus.
With the same aim we show in Fig.\ \ref{MC2-p-pi0}
a two-dimensional angular distribution of energy depositions
$dE/dz\,d\phi$ for photons, decay products of $\pi^0$.

\begin{figure}[hbt]
\centerline{
  \includegraphics[width=0.9\textwidth, bb= 15 170 550 630]{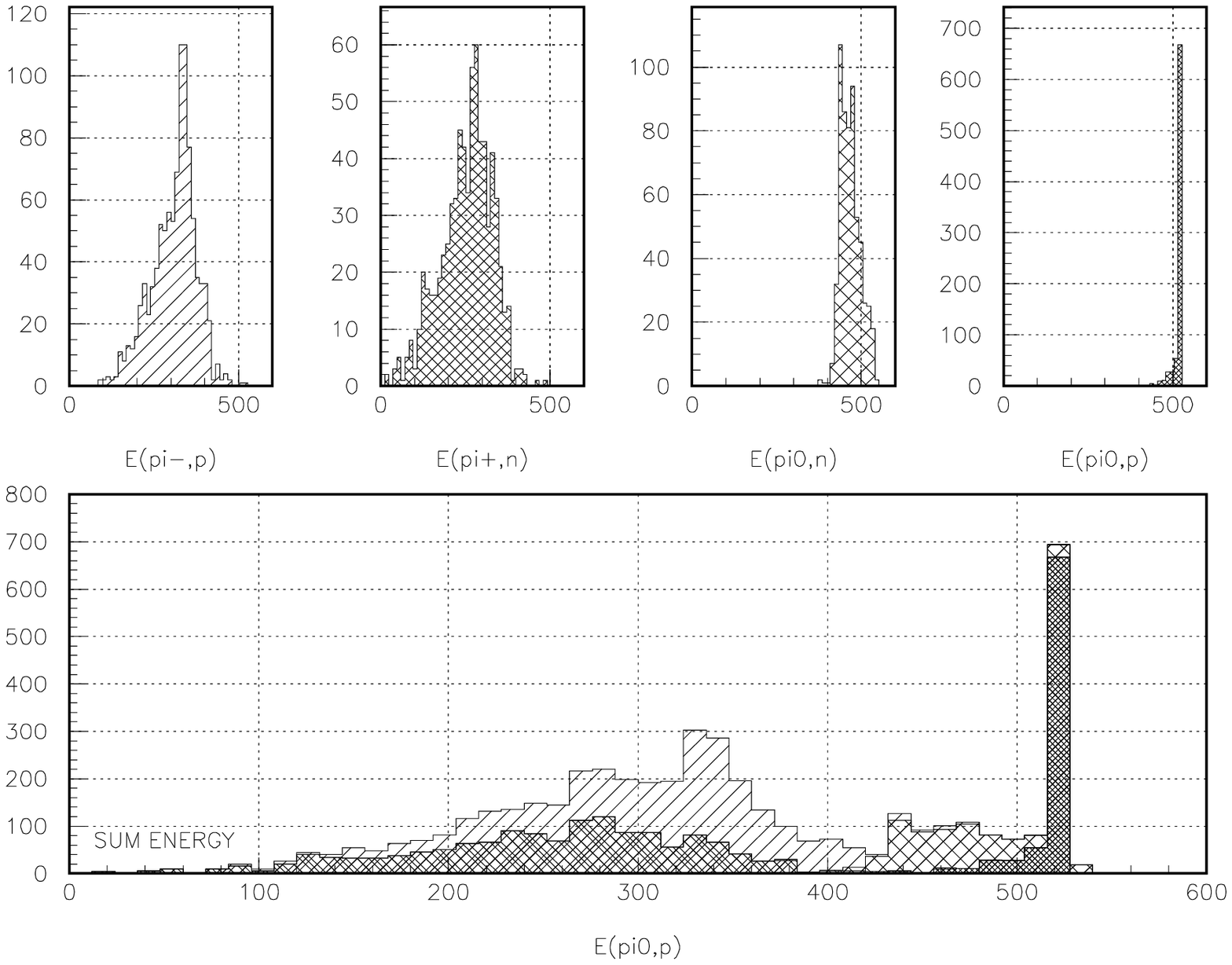}}
\caption{Top: energy depositions of $\pi N$ pairs with different
charges in the BGO calorimeter. Bottom: a combined (with appropriate
weights 1/3, 1/3, 1/6, 1/6) distribution which thus gives an idea of
a background.}
\label{fig:graad2}
\end{figure}

In Figure \ref{fig:graad2}, shown are simulated distributions over the
total energy deposited by $\pi N$ pairs of different electric charges in
the model BGO detector.
Then all four decay channels of
$\eta$-nuclei, i.e.\ $\pi^+ n$, $\pi^- p$, $\pi^0 p$ and $\pi^0 n$, are
added with appropriated weights (1/3, 1/3, 1/6 and 1/6, respectively) to
give a combined histogram shown in the bottom of Fig.~\ref{fig:graad2}.

This figure clearly shows that the calorimetric method of detection of
the $\pi N$ pairs is quite suitable only for the $\pi^0 p$ channel.
Three other channels do not assure a full energy deposition and lead to
large fluctuations. They rather give a broad background which though can
be suppressed by veto detectors or alike. Note that there are also other
contributions to the background, e.g.\ one from double-pion
photoproduction off the nucleus. Moreover, energy losses in a
bigger target will be larger.

\begin{figure}[hbt]
\centerline{
  \includegraphics[width=0.9\textwidth, bb= 15 170 550 630]{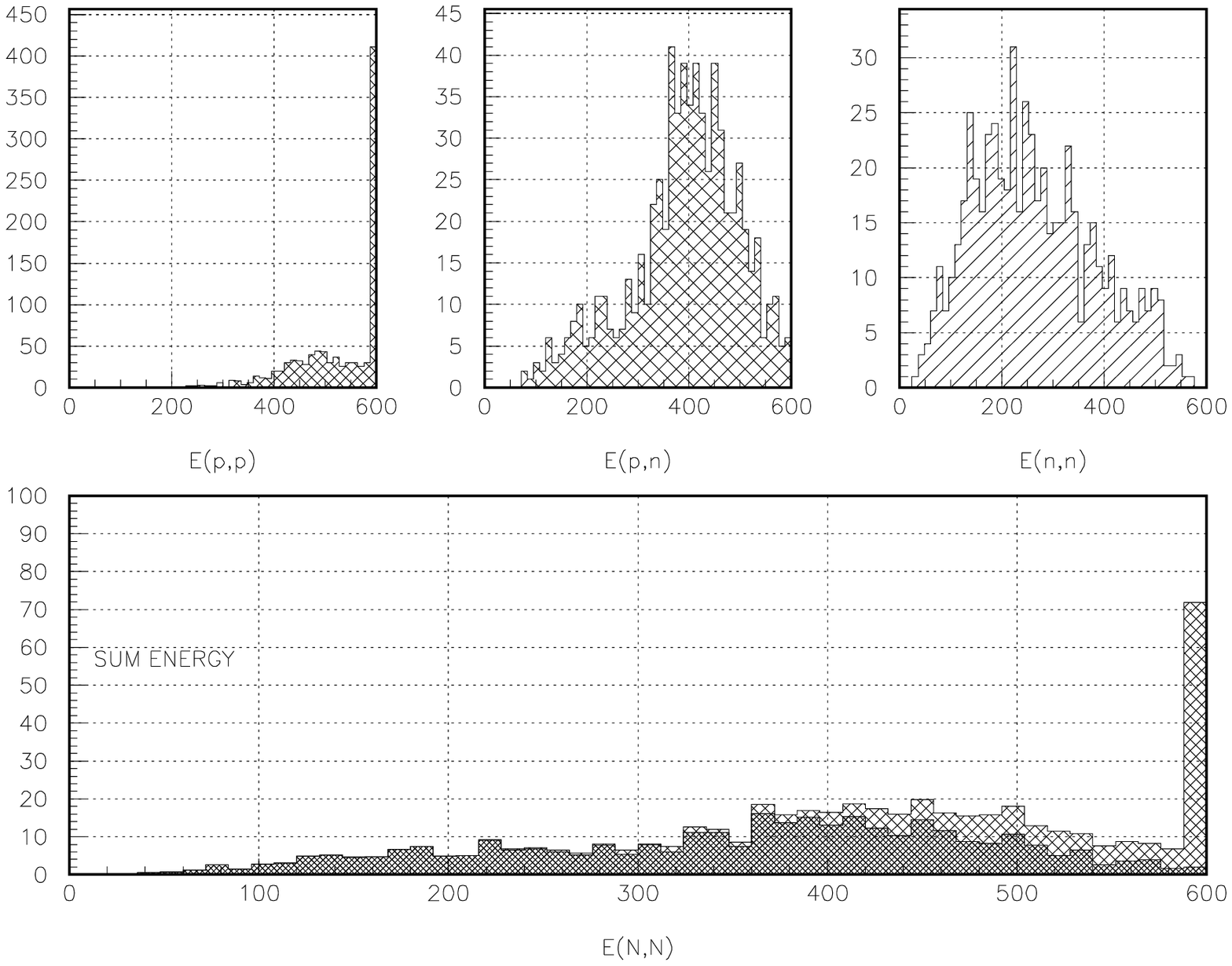}}
\caption{Top: energy depositions of $NN$ pairs with different
charges in the BGO calorimeter. Bottom: a combined distribution
(with appropriate weights 1/3, 1/3, 1/6, 1/6); it gives an idea of
background.}
\label{fig:ppw2}
\end{figure}

From all these figures we conclude that among 4 charge states $\pi N$
only the channel $\pi^0p$ can be used for accurate measurements
suitable for the purposes of the present experiment.
The channel $pp$ is suitable too, as can be inferred from
Fig.\ \ref{fig:ppw2}.
Thus, together with the forward-flying proton ($E_{p_1}\sim 250$ MeV),
we choose the coincidence $(p_1,\gamma\gamma[\pi^0],p)$
as the trigger for the reaction
(\ref{eq:reac1}), Fig.~\ref{reac1}, where $\gamma\gamma[\pi^0]$ means
two photons with the invariant mass of the neutral pion.
The trigger for the reaction (\ref{eq:reac2}), Fig.~\ref{reac1},
is the triple coincidence $(p_1,p_2 p_3)$ of the
forward-flying proton with two protons in the BGO calorimeter.
At last, the trigger for the reaction (\ref{eq:reac3}), Fig.~\ref{reac3},
is the coincidence $(p_1,\gamma\gamma[\eta])$, where $\gamma\gamma[\eta]$
means two photons with the invariant mass of the $\eta$-meson;
though, six-photon events from the $\eta\to 3\pi^0$ could also be recorded
and then analyzed.
Note that $\pi^0p$ pairs produced through annihilation of slow $\eta$
in the nucleus have a nearly isotropic distribution and have the opening angle
$\theta_{\pi^0 p}$ close to $180^\circ$. In the case of the two-photon decay
of slow $\eta$  (or the two-proton decay of $S_{11}(1535)$ through
$S^{+}_{11}(1535)+p \to p_2+p_3$)
the angular distribution of photons (protons) is again nearly
isotropic and the opening angle $\theta_{\gamma_1\gamma_2}$
($\theta_{p_2 p_3}$) is close to $180^\circ$.

\begin{figure}[hbt]
\centerline{
\includegraphics*[height=6cm]{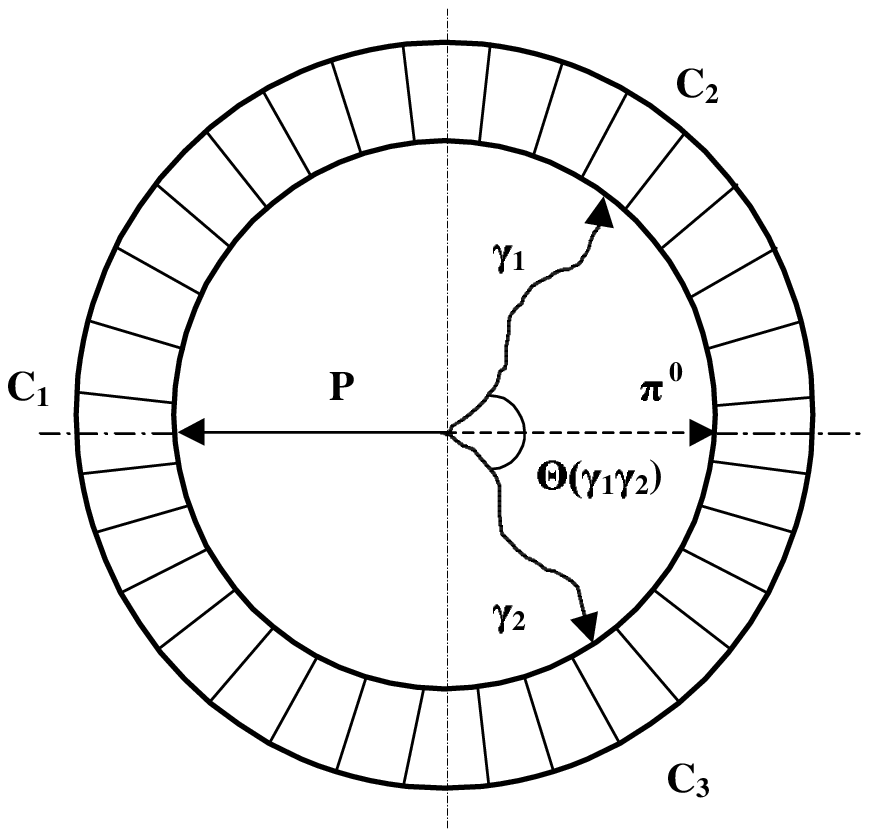}\qquad
\includegraphics*[height=6cm]{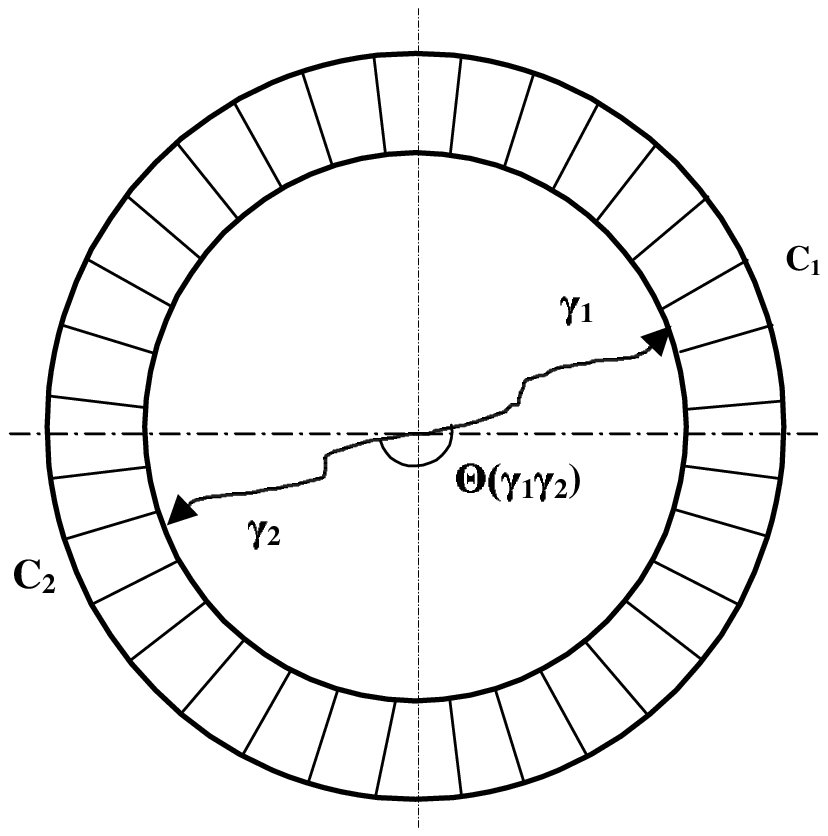}
}
\caption{The topography of detected events in the BGO calorimeter.}
\label{topo-events}
\end{figure}

The electronic trigger of the events in the $(p_1,\pi^0p)$ channel
will include simultaneous signals in the cylindrical wire chambers
and in three BGO counters $C_1$, $C_2$, $C_3$ (see Fig.\ \ref{topo-events}).
The signals in the wire chambers show the path of the charged particle
(the proton), with the $C_1$ BGO-counter giving the energy loss
of the proton, $\Delta E_p$.
Signals in the $C_2$ and $C_3$ counters lying opposite to the $C_1$ counter
indicate a detection of two photons from the $\pi^0$ decay.
The total energy deposited in the
BGO crystals plus the energy carried by the forward-flying proton $p_1$
must match the initial photon energy $E_\gamma$ up to, possibly,
a small excitation energy of the rest of the nucleus.

\subsection{Estimates of count rates}

$\bullet$
The count rate of $\eta$-nuclei production through the coincidences
$(p_1, \pi^0 p)$ at GRAAL can roughly be estimated as
\begin{equation}
     Y(p_1, \pi^0 p) = \sigma({}_\eta A) N_\gamma N_n
      \frac{\Delta\Omega_p}{4\pi}
      f_{\pi^0} f_{p_1} {\rm Br}(\pi^0 p)
  \simeq 500 ~\rm \frac{events}{day},
\label{yield1}
\end{equation}
where

$\sigma_t({}_\eta A) \simeq 0.5 ~\mu$b is the total cross section
of $\eta$-nuclei photoproduction inside a nucleus
in the reaction  $\gamma + {}^{12}{\rm C}\to p + {}^{11}_{~\eta}{\rm B}$
\cite{try95}, in which only nuclear protons contribute to $\eta$ production;
an average over the energy interval of $E_\gamma = 700{-}1000$ MeV is assumed;

$N_\gamma \simeq 2\cdot 10^5~{\rm s}^{-1}$ is the number of tagged photons
over the energy interval $700{-}1000$ MeV;

$N_n \simeq 7\cdot 10^{23}$ is the number of nuclei $^{12}$C in the target
of the length 6 cm;

$\Omega_p \simeq 11 ~\rm rad$ is the solid angle covered by
the BGO spectrometer where protons are detected;

$f_{\pi^0} \simeq 0.8$ is the efficiency of
detecting $\gamma_1\gamma_2$ from the $\pi^0$ decay in the BGO spectrometer;

$f_{p_1} \simeq 0.7$ is a fraction of forward-flying protons
generated in the process of $\eta$-nucleus formation
and detected by the time-of-flight wall;

${\rm Br}(\pi^0 p) = \frac{1}{6} $ is the branching ratio to generate the
channel $\pi^0p$ in the process of annihilation of $\eta$ inside the nucleus
among four possible alternatives $\pi^0p$, $\pi^+n$, $\pi^-p$, $\pi^0n$.

The count rate (\ref{yield1}) is quite reasonable for observation
because of a strong suppression of backgrounds due to multiple coincidences.

$\bullet$
The count rate of $\eta$-nuclei production through the coincidences
$(p_1, p_2 p_3)$ at GRAAL can roughly be estimated as
\begin{equation}
     Y(p_1, p_2 p_3) = \sigma({}_\eta A) N_\gamma N_n
      \Big(\frac{\Delta\Omega_{p_2}}{4\pi} \Big)
      f_{p_2p_3} {\rm Br}(pp)
  \simeq 120 ~\rm \frac{events}{day},
\label{yield2}
\end{equation}
where

$f_{p_2p_3} \simeq 0.8$ is an efficiency to detect the correlated
proton pair from the subprocess $S_{11}^+(1535) + p\to p_2 p_3$;

${\rm Br}(pp) \simeq 0.1 \cdot \frac {1}{4}$
is the branching ratio of the $\eta$-nucleus
decay through the two-nucleon absorption (about 10\% of $NN$ pairs
in comparison with 90\% of $\eta N$ decays through the $\pi N$ channel
\cite{chi91}) times the fraction 1/4 to have $NN=pp$.

$\bullet$
The count rate of slow-$\eta$ production through the coincidences
$(p_1,\gamma_1\gamma_2[\eta])$ at GRAAL can roughly be estimated as
\begin{equation}
     Y(p_1, \eta) = \sigma_t(\eta) N_\gamma N_n f_{\eta} f_{p_1}
  \simeq 75000 ~\rm \frac{events}{day},
\label{yield3}
\end{equation}
where

$\sigma_t(\eta) \simeq 25\mu$b \cite{kru96} is the total cross section
of $\eta$ photoproduction off protons in the nucleus $^{12}$C
with the kinetic energy of $\eta$ below 70 MeV;
this condition corresponds to the backward-angle production of $\eta$
in the subprocess $\gamma p\to\eta p$;
an average over the energy interval of $E_\gamma = 700{-}1000$ MeV
is assumed;

$f_{\eta} \simeq 0.8$ is the efficiency of
detecting $\gamma_1\gamma_2$ from the $\eta$ decay by the BGO spectrometer;

$f_{p_1} \simeq 0.3$ is a fraction of forward-flying protons
generated in the process of slow-$\eta$ production
and detected by the time-of-flight wall.

Note that the count rate of the $(p_1, \gamma_1\gamma_2[\eta])$ events coming
from radiative decays $\eta\to\gamma_1\gamma_2$ of the bound
$\eta$ is expected to be $\sim 10^4$ times less than
Eq.\ (\ref{yield1}) gives, thus being of order $0.1$ events/day and too small
to be detected.

\section{Summary of aims and perspectives}

In conclusion, studies of $\eta$-mesic nuclei lie at the interception
of the nuclear physics and the physics of hadrons and they
promise to bring a new information important for both the fields. Such
studies are quite feasible at electron accelerators with
bremsstrahlung beams. Tagged photon beams can also be used.
The main aims and perspectives of the proposed experiment
can be summarized as follows:

\begin{itemize}

\item
Generally, we expect to bring a novel experimental information on behavior
of the $\eta$-meson in the nuclear matter.  Specifically, we hope to
measure energy levels $E_\eta$ and widths $\Gamma_\eta$ of the bound
$\eta$-nucleus systems and compare them with modern calculations.
It is worth to say that theoretical values for $E_\eta$ and $\Gamma_\eta$
strongly depend on the potential $V_{\eta N}$ and the scattering amplitude
$f_{\eta N}$ assumed for the elementary $\eta N$ interaction
and on the dynamics of dressing $\eta$ and $S_{11}(1535)$
in the nuclear matter \cite{liu86,kul98,gar02,chi91,jid02,hai02}.

\item
We will measure separately two different channels, $\pi N$ and $NN$,
of $\eta$-nucleus decays. This will provide a valuable test for dynamics
of the $S_{11}(1535)$ resonance in the nuclear matter.

\item
In parallel, we will study near-threshold production of $\eta$-mesons which
is expected to have a strong energy dependence due to attractive forces
between $\eta$ and the nucleus. The data on this energy dependence provide
a further useful input for constraining existing theoretical models for
$\eta N$ and $\eta A$ interaction.

\item
In a perspective, studies of $\eta$-nuclei can open a way towards
wider investigations of $(\rho, \omega, \varphi)$-nuclear systems
which are presently discussed in the literature \cite{tsu98}.

\end{itemize}

\acknowledgments

This work was supported by the Russian Foundation for Basic Research, grant
\#02-02-16519.

\end{document}